\documentclass[12pt,preprint]{aastex}
\usepackage{graphicx}
\usepackage{epsfig}
\usepackage{times}
\usepackage{natbib}
\usepackage{url}
\usepackage{amssymb,amsmath}
\usepackage{hyperref}
\usepackage[usenames,dvipsnames]{color} % For color text: \color command
\usepackage[normalem]{ulem}     %Provides the following commands:
\newcommand{\fig}[1]{Figure~\ref{fig_#1}}
\newcommand{\figS}[1]{Figures~\ref{fig_#1}}
\newcommand{\figs}[2]{Figures~\ref{fig_#1} and \ref{fig_#2}}

\newcommand{\sect}[1]{Section~\ref{sect_#1}}

\newcommand{\eq}[1]{Equation~(\ref{eq_#1})}

\newcommand{\BE}{\begin{equation}}
\newcommand{\EE}{\end{equation}}
\newcommand{\BA}{\begin{eqnarray}}
\newcommand{\EA}{\end{eqnarray}}

\newcommand{\degree}{^\circ} 
\newcommand{\kms}{km s$^{-1}$} %{km/s} 
\newcommand{\Rsun}{$R_{\odot}$}  

\newcommand{\eg}{\textit{e.g.,}}

\newcommand{\ha}{H$\alpha$}

\shorttitle{EUV Loop Contraction Followed by Expansion}
\shortauthors{Chandra et al.}

\begin{document}

\title{Filament Eruption Driving EUV Loop Contraction then Expansion above a Stable Filament} 
\author{Ramesh Chandra\altaffilmark{1}, Pascal D\'emoulin\altaffilmark{2, 3}, Pooja Devi\altaffilmark{1},
Reetika Joshi\altaffilmark{1}, Brigitte Schmieder\altaffilmark{2, 4, 5}}

\affil{$^1$ Department of Physics, DSB Campus, Kumaun University, Nainital -- 263 001, India \email{rchandra.ntl@gmail.com}}
\affil{$^2$ LESIA, Observatoire de Paris, Universit\'e PSL, CNRS, Sorbonne Universit\'e, Universit\'e de Paris, 5 place Jules Janssen, 92195 Meudon, France }
\affil{$^3$ Laboratoire Cogitamus, 1 3/4 rue Descartes, 75005 Paris, France}
\affil{$^4$ Centre for Mathematical Plasma Astrophysics, Department of Mathematics, KU Leuven, 3001 Leuven, Belgium}
\affil{$^5$ LSUPA, School of Physics and Astronomy, University of Glasgow, Scotland}

%*****************************************************************************
\begin{abstract}
We analyze the observations of EUV loop evolution associated with the filament eruption located at the border 
of an active region. The event SOL2013-03-16T14:00 was observed with a large difference of view point by the 
Solar Dynamics Observatory and Solar Terrestrial Relations Observatory --A spacecraft. The filament height 
is fitted with the sum of a linear and exponential function. These two phases point to different physical mechanisms 
such as:  tether-cutting reconnection and a magnetic instability. While no X-ray emission is reported, this event 
presents the classical eruption features like: separation of double ribbons and the growth of flare loops. 
We report the migration of the southern foot of the erupting filament flux rope due to the interchange reconnection 
with encountered magnetic loops of a  neighbouring AR. Parallel to the erupting filament, a stable filament 
remains in the core of active region. The specificity of this eruption is that coronal loops, located above the 
nearly joining ends of the two filaments, first contract in phase, then expand and reach a new stable configuration 
close to the one present at the eruption onset. Both contraction and expansion phases last around 20 min. The main 
difference with previous cases is that the PIL bent about 180$^0$ around the end of the erupting filament 
because the magnetic configuration is at least tri-polar. These observations are challenging for models which 
interpreted previous  cases of loop contraction within a bipolar configuration.  New simulations are required to broaden the 
complexity of the configurations studied.
\end{abstract}
%*****************************************************************************

%%%%%%%%%%%%%%%%%%%%%%%%%%%%%%%%%%%%%%%%%%%%%%%%%%%%%%%%%%%%%%%%%%%%%%%%%%%%%%%%%%%%%%
\section{Introduction}
\label{sect_Introduction}
%%%%%%%%%%%%%%%%%%%%%%%%%%%%%%%%%%%%%%%%%%%%%%%%%%%%%%%%%%%%%%%%%%%%%%%%%%%%%%%%%%%%%%
 
%{\S}{\bf --- Filament/Prominence} \\ 
Solar filaments (or prominences, when observed at solar limb) are cool and dense material plasma suspended in a million degree hot corona \citep{Mackay2010, Labrosse2010, Parenti2014}.  They are located above the photospheric inversion line (PIL) of the vertical component of the magnetic field.  They are observed in the active as well as in the quiet solar atmosphere or between two active regions.  Since the plasma of filaments is about a factor one hundred denser than the coronal plasma it needs to be supported by a force against the action of gravity.   
The existence of stable support in magnetic dips was initially proposed by \citet{Kippenhahn1957}. Present models typically involved a magnetic structure with the filament plasma caught in magnetic dips \citep[see the review by][]{Mackay2010}.
Magnetic dips can be present naturally in potential fields with a quadrupolar magnetic configuration but not with a bipolar configuration.  Several models have been developed such as the sheared arcade model \citep{Antiochos1994}, and the flux rope (FR) model \citep{Aulanier1998, Priest1989, Aulanier2002}.  Finally, the existence and the evolution of a filament is strongly linked to the associated magnetic field configuration.

%{\S}{\bf --- Filament eruptions} \\
Usually, at some point of their evolution filaments become unstable and erupt.  
The eruption can be confined by the overlying magnetic field, then is called a failed eruption.  
In the opposite case, the filament, the surrounding coronal magnetic field and the plasma become a coronal mass 
ejection (CME) which is ejected towards the interplanetary medium \citep[\eg\ ][]{Gibson2006, Chandra2010, Schmieder2013, Kliem2014}. Therefore filament eruptions are thought to play a diagnostic role for the origin of CMEs 
because of their much higher plasma density than the surrounding corona.
Eruptions and associated phenomena, such as the formation of flare ribbons and their separation with time, were initially explained by the CSHKP model \citep{Carmichael1964, Sturrock1966, Hirayama1974, Kopp1976}.  
Later on,  this 2--dimension model was extended into a 3--dimensional model 
that could explain eruption associated phenomena including slipping 
reconnection, circular ribbon formation, and the magnetic shear evolution of 
flare loops \citep{Aulanier2010, Janvier2015}. 
 
Different mechanisms  could explain the trigger of eruptions, which include 
the magnetic breakout \citep{Antiochos1999}, tether cutting \citep{Moore2006}, kink 
instability \citep{Torok2005} and the torus instability or catastrophe model \citep{Kliem2006}. 
The observations and models of both the possible pre-eruptive magnetic configurations and the eruption mechanisms have recently been reviewed \citep{Green2018,Georgoulis2019,Patsourakos2020}.
As the FR is erupting according to the  \citet{Lin2000} model, a thin current sheet (CS) is formed behind the erupting FR. The erupting FR can be ejected from the corona into the heliosphere thanks to the magnetic reconnection occurring in the current sheet behind the FR. This reconnection transforms part of the stabilizing magnetic connections passing above the FR to connections located below the FR (flare loops) and to connections wrapped around the FR, so further building the FR.

%{\S}{\bf --- Loop Contraction/Expansion (Observations)} \\ 
One of the remarkable phenomena accompanying solar eruptions is the evolution of coronal loops. 
This phenomenon is reported with various space borne observations in flare loops with the shrinkage of individual loops while the global loop system expand upwards as further flare loops are formed \citep{Forbes1996, Li2006, Sui2004, Zhou2008, Joshi2009}. 
Moreover, since the launch of {\emph SDO} in 2010, observational evidences of coronal loop contraction and expansion, not associated to flare loops, are increasing \citep{Sun2012, Gosain2012, Liu2012, Zhou2013, Simoes2013, Shen2014, Dudik2017, Wang2018, Dudik2019, Devi2021}.  This is a consequence of {\emph SDO} continuous high temporal and spatial resolution observations.  This phenomena is usually observed together with eruptive solar flares.  These loops are already present before the flare, so they are different from flare loops which are formed by reconnection during the flare.  The speed of the contraction and expansion is typically between a few \kms ~and~ 40 \kms. Furthermore, in some cases the contracted loops end up oscillating \citep[\eg\ ][]{Gosain2012, Liu2012, Simoes2013}.

%{\S}{\bf --- Existing models} \\ 
To explain the phenomena of the loop contraction two mechanisms were explored. 
The first mechanism is the ``magnetic implosion'' conjecture proposed by \citet{Hudson2000}. 
According to this the energy generated during a solar eruption has its origin in 
the implosion of the magnetic configuration in a nearby part of the corona. 
The expansion followed by contraction in coronal loops is recently simulated in 3D by \citet{Wang2021}.
 The physical interpretation of such implosion has later changed.  For example, \citet{Simoes2013} interpret the observed 
loop contraction with an implosion which is a consequence, and not a driver, of the energy release during the event.  
Later on, \citet{Russell2015} interpret the loop contraction as the consequence of the reduction of the magnetic pressure due to the magnetic reconnection occurring in the flare.   
Several observations of loop contraction are interpreted as supporting this mechanism \citep{Wang2018}.
The second mechanism is proposed by \citet{Aulanier2010} and \citet{Zuccarello2017}.  This model is based on the 3D magnetohydrodynamics (MHD) numerical simulations of the formation then eruption of a FR in a bipolar magnetic configuration modelling an AR configuration.  According to their model, the loop contraction/expansion is the result of the vortex forming on both sides of erupting FR legs.  These vortex flows drive the coronal loops and as a result of this the contraction/expansion in loops observed, depending where the loops are in the vortex. The careful analysis of the contraction/expansion of loops in some studied eruptions filament support the conclusion of the MHD simulations \citep{Dudik2017, Dudik2019, Devi2021}.

 Coronal loops can also be excited by propagating coronal waves, in particular fast magneto-acoustic 
waves \citep{Wills1999, Ballai2007, Ballai2008}. These waves can be generated by sudden energy 
release process like the phenomena of solar flares and CMEs. The propagation of coronal waves can be considered as a freely propagating  wavefront which is observed to interact with coronal loops \citep[see, e.g.][] {Wills1999,Ballai2008}.   %P

%{\S}{\bf --- Aim of the paper} \\
In this paper, we study loop contraction and expansion associated with a filament eruption on 2013 March 16 observed in EUV wavelengths by the AIA imager aboard SDO and by EUVI aboard STEREO with a large difference of viewing points. 
The specificity of this event is that the loop contraction and expansion were occurring not only in coronal loops located nearby the end point of the erupting filament, but also in loops located over a stable filament.  
This filament is located on the side of the erupting one with a nearly parallel orientation of 
most parts of both filaments, which are also nearly joining at one end. Then, this is an 
interesting eruption to study as it broaden the range of configurations where 
the  loop contraction and expansion could occur.   
The layout of the paper is as follows: \sect{Observations} describes the data sets, the 
morphology of filament eruption, the history of the  magnetic configuration of the involved active region (AR), the kinematics of the eruption, and the observational results related to EUV loop contraction and expansion.  The physical interpretation of the obtained results is given 
in \sect{Physical}.  Finally in \sect{Conclusion}, the conclusion of the study is given.

%#########################################################
\begin{figure*}    % Figure 1
\centering
\includegraphics[width=0.8\textwidth]{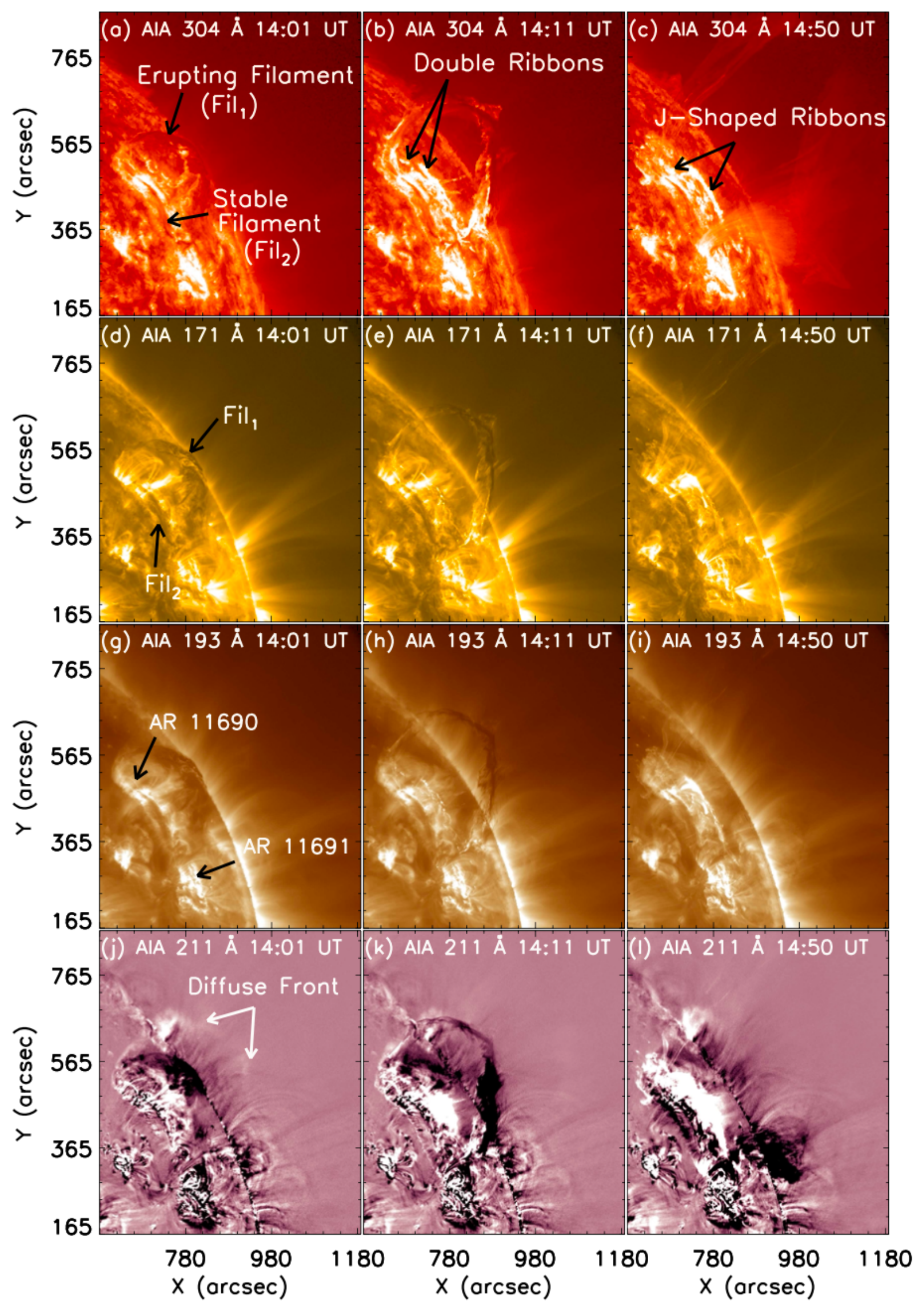}
\caption{Multi-wavelength view of the filament eruption of 2013 March 16 in 
AIA 304, 171, 193 { and 211 \AA .  The AIA 211 \AA\ images are base difference images (base time: 12:30 UT). }
The location of the erupting filament (Fil$_1$), the stable filament (Fil$_2$)  
and the flare ribbons are indicated in panels (a), (d) and (b, c), respectively. 
The AR involved in the eruption, AR 11690, and the southern one are indicated in panel (g). 
{ The propagating diffuse front above the filament is shown in panel (j).}
Movie of these data is available in the Electronic Supplementary Materials. The animation starts at 12:30 UT and end at 15:00 UT. The realtime duration of the animation is 25 seconds.
} 
\label{fig_evolution}
\end{figure*}
%#########################################################

%#########################################################
\begin{figure*}    % Figure 2
\centering
\includegraphics[width=\textwidth]{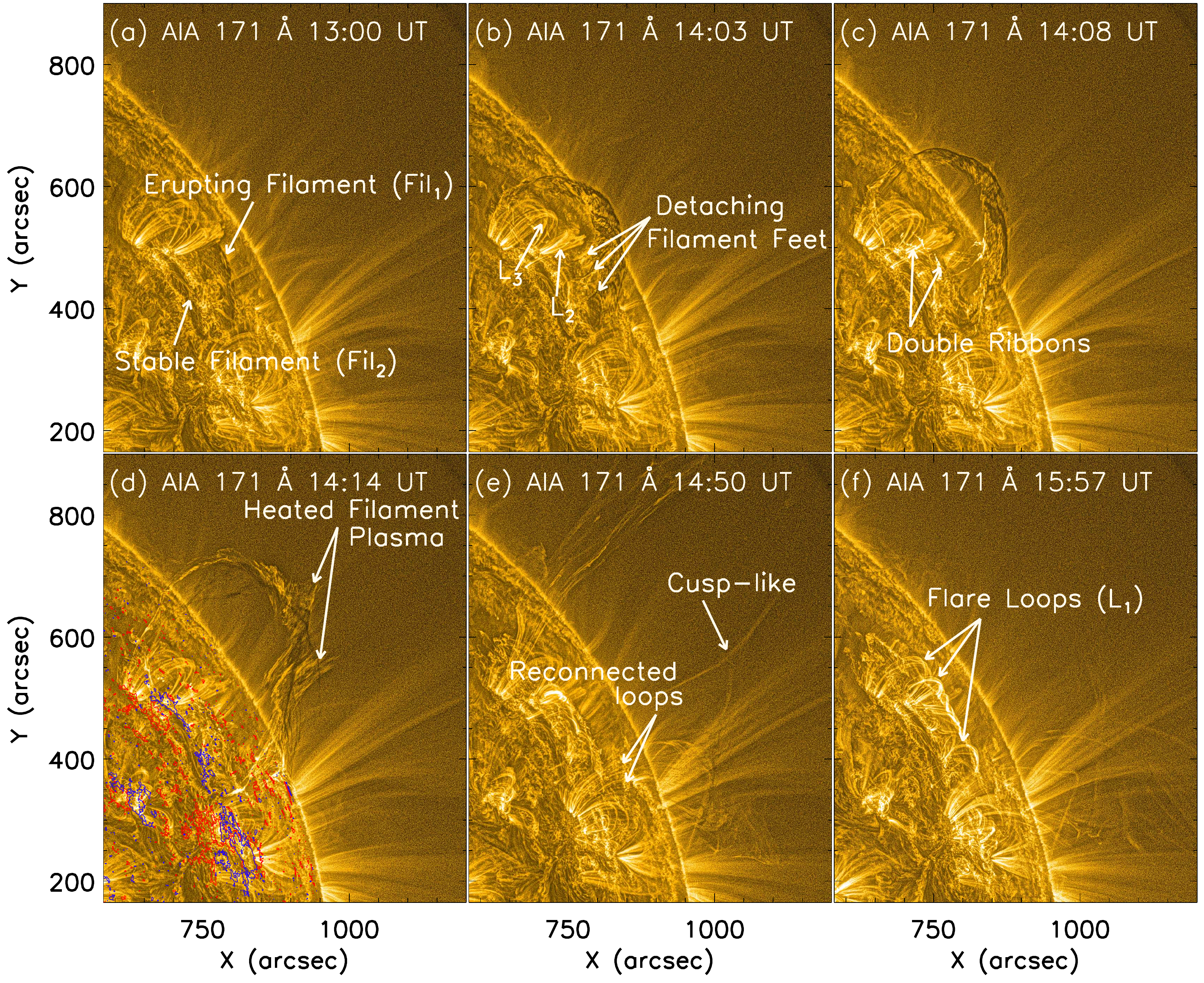}
\caption{Evolution of the filament eruption with AIA images at 171 \AA\ processed with the MGN method (\sect{Overview}). The erupting (Fil$_1$) and the stable (Fil$_2$) filaments are indicated in panel (a). Specific characteristics of the erupting filament are pointed in panels (b) and (d).  The consequences of the main energy release are indicated in panels (c) and (f) with flare ribbons and flare loops (L$_1$), respectively.   
The HMI positive/negative polarity contours are overlaid in panel (d) with red/blue 
contours (contour level: $\pm 30$ G).
The loop systems L$_2$ and L$_3$, located above Fil$_2$, are indicated in panel (b). 
The reconnected loops, marked in panel (e), are due to the reconnection of the southern 
leg of the erupting FR with the magnetic field of AR 11691 (located southward of AR 11690, see \fig{evolution}g). 
The associated movie is available in the Electronic Supplementary Materials. The animation starts at 12:30 UT and end at 15:00 UT. The realtime duration of the animation is 25 seconds. 
}
\label{fig_MGN171}
\end{figure*}

\section{Observations}
\label{sect_Observations}

\subsection{Overview of the Event}
\label{sect_Overview}

%   {\S}{\bf --- Instruments} \\      
The filament eruption on 2013 March 16 was observed by the \emph{Atmospheric Imaging Assembly} \citep[AIA,][]{Lemen2012} onboard \emph{Solar Dynamics Observatory} \citep[SDO,][]{Pesnell2012} at different 
EUV wavelengths with pixels of 0.6 $\arcsec$ and cadence of 12 s. 
For our present analysis, we  used the AIA data of 304, 171, 193 {  and 211 \AA\ } which correspond to the wavelengths where the eruption is  observed {the best}.  For the magnetic configuration of the filament region, we analyzed the photospheric magnetic field line-of-sight data of \emph{Helioseismic Magnetic Imager} \citep[HMI,][]{Schou2012}. The pixel resolution and the temporal resolution of the HMI magnetic field data is 0.5 $\arcsec$ and 45 s respectively.
The eruption was also observed by the \emph{Solar Terrestrial Relations Observatory Ahead} \citep[STEREO~A,][]{Howard2008} spacecraft with a cadence of 10 min.  
All the above mentioned data are processed by the solar soft package. 
To remove the solar rotation, we  rotated all used  AIA analyzed images at 12:30 UT. 
For this purpose, we  used the ``drot$\_$map'' routine available in solar soft.
In order to enhance the contrast of the AIA data, we processed the images with the Multi-Gaussian Normalization (MGN) method developed by \citet{Morgan2014}.
To further outline the evolution we create base-difference movies, both with and without MGN method, with the base time set at 12:30 UT.     

%   {\S}{\bf --- Global description} \\
The erupting filament (Fil$_1$) on 2013 March 16 was located at the north-west border of the NOAA AR 11690 (\fig{evolution}g).
A second stable filament (Fil$_2$) is present on the main PIL of AR 11690.
Both filaments are indicated with arrows in \fig{evolution}a,d.
The evolution of the erupting filament Fil$_1$ in AIA 304, 171, 193 and 211 \AA\ is summarized with three selected times in \fig{evolution}. The accompanying movies provide a detailed view of the eruption. The filament started to move upwards at $\approx$ 13:19 UT in the north-west direction (best seen with base-difference movies). 
Much later on, at about 13:59 UT, two flare ribbons started to develop below the erupting filament.  The double ribbons, indicated by arrows in \fig{evolution}b, are well observed in 304, 171, 193, and 211 \AA . As the filament was moving upward, the double ribbons separated from each other as predicted in the standard flare model (\sect{Introduction}). Furthermore, J-shaped ribbons are present after 14:29 UT which indicate the eruption of a FR \citep{Demoulin1996,Aulanier2010}. These ribbons are best seen in the base-difference of AIA 304 \AA .

%   {\S}{\bf --- Best observations of eruption} \\
The MGN technique is applied to AIA data in order to enhance the coronal structures so that the morphology and the dynamics of the eruption could be better understood.  \fig{MGN171} shows the evolution with the MGN technique applied to 171 \AA . Arrows point to the main observational features.  The filament eruption is observed the best with the 171 \AA\ filter with more contrasted images which reveal both the EUV absorption by the dense and cold filament plasma and the emission of heated plasma (\eg\ part of the filament, flare ribbons and loops).  

 On mid 2013 March 16 the filament Fil$_1$ erupted as a coherent structure. 
This is emphasized by a sharp leading edge moving as a coherent entity (Figure 2a-c). 
At the beginning of the filament eruption three dark features are extending below the filament body. They are pointed by three arrows in Figure 2b. These extensions are filament feet/barbs which have been identified previously mostly in quiescent filaments \citep{Aulanier1999, Mackay2010}. 
As the eruption progresses, the feet of the filaments become more clearly visible, before getting split with the lower part falling towards the chromosphere and with the upper part being integrated in the eruptive configuration. 
During this splitting, the associated plasma is changing from absorption to emission, so it is heated. 
They are both indications that magnetic reconnection occurs below the erupting FR.

As the filament erupts, its two ends stay anchored at low heights (\fig{MGN171}a-d). This allows  to identify the 
two footpoints of the erupting FR which are located in the northward and west-southward periphery of AR 11690, in negative and positive photospheric polarities, respectively (Figure 3d).  Later on in the eruption, a cusp-like 
shape appears in emission in 171 \AA\ (\fig{MGN171}e).  However, the temporal evolution is more compatible with the crossing in projection of two loops (see associated movie). 

\begin{figure*}    % Figure 3
\centering
\includegraphics[width=0.9\textwidth]{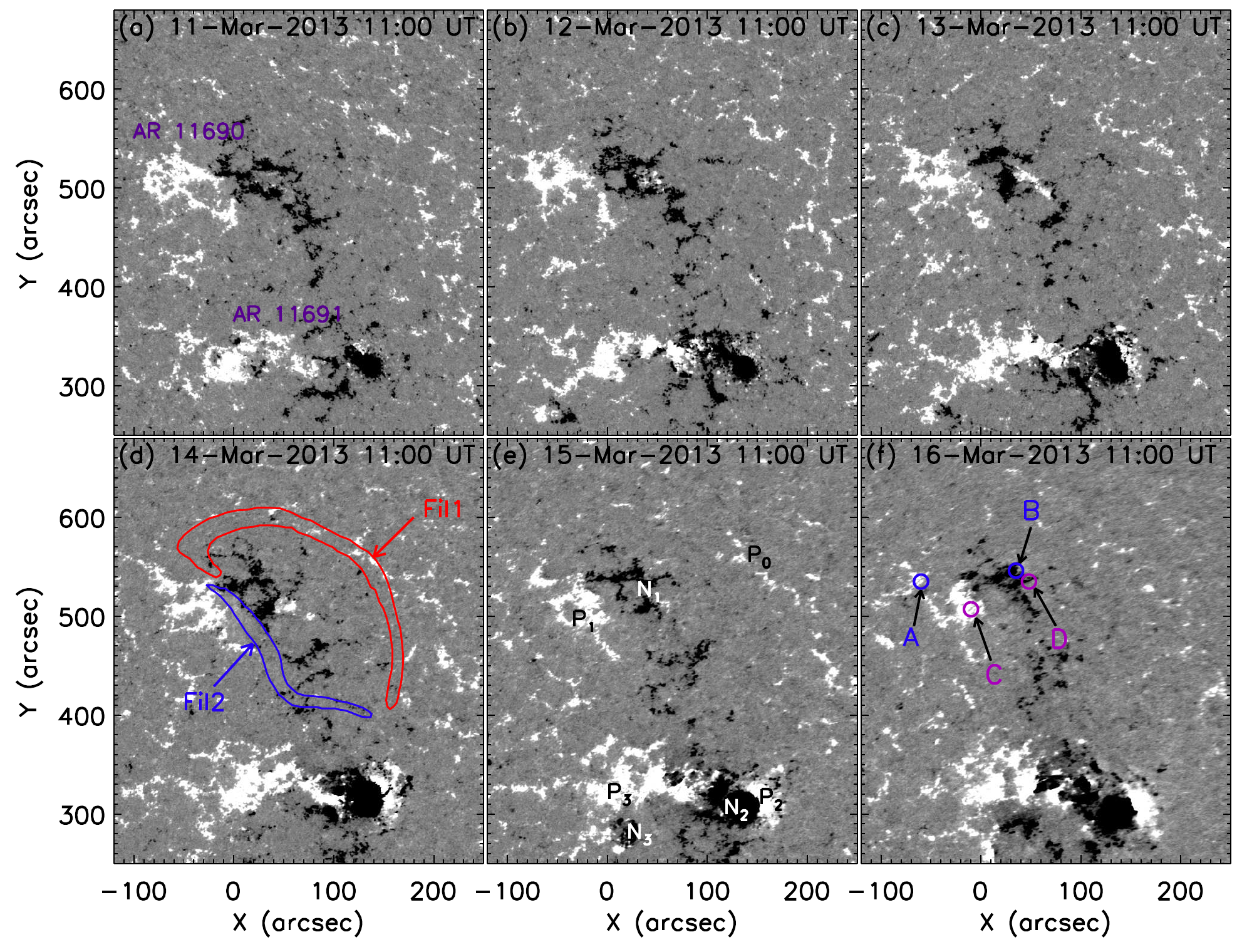}
\caption{Evolution of the longitudinal component of the photospheric magnetic field the ARs 11690 and 11691. The observations are derotated to the 
central meridian location.  The contours in panel (d) delimitate the location of the two 
filaments observed in \ha . The red contour indicates the erupting filament Fil$_1$, while 
the blue contour indicates the stable filament Fil$_2$. The different magnetic polarities 
are named by $P_0$ (remnant field  of a previous AR), $P_1, N_1$ (AR 11690), $P_2, N_2, P_3, N_3$ (AR 11691). 
The locations of the loop foot-points, defined in \fig{loops}b, are indicated by the pairs A, B and C, D, 
respectively in panel (f).   An associated movie from 2013 March 11 00:00 UT to 2013 March 16 17:00 UT of realtime duration 22 seconds is available in the Electronic Supplementary Materials.
} 
\label{fig_mag}
\end{figure*}

The evolution  of the southern part of the erupting filament is complex (see attached movies in 304, 171, 193 
and 211 \AA). We interpret this evolution as the reconfiguration of the southern FR leg.  During its 
upward ejection, the FR expands and reconnects, at least partially, with the main bipolar field of AR 11691.  
This shifts the FR magnetic anchorage from the positive  polarity in front of AR 11690 to the positive following polarity of AR 11691 as traced by the new connections pointed in \fig{MGN171}e (with ``Reconnected Loops'').
 A part of the dense filament plasma fall along and fill partly these new connections as seen the best in 171 and 304 \AA\ base-difference movies. 
We conclude that the lateral drift of the filament foot is accomplished via interchange reconnection with encountered southern magnetic loops. These observations are globally consistence with the 
previously reported observations of such lateral drift of the foot of an erupted prominence/filament
\citep{Hori2000, vanDriel-Gesztelyi2014, Dudik2019, Lorincik2019, Zemanova2019}.
{ Another part of the erupted filament drains to the end region of the J-shaped flare ribbon in the west (towards polarity $P_0$ in \fig{mag}), which is near the periphery of the flux rope leg. This is best seen in AIA higher temperature, e.g. 211 and 193 \AA , and it implies that the magnetic configuration of this filament leg splits.}

%   {\S}{\bf --- Flare} \\
 The flare associated to this filament eruption is weak as it does not saturate the EUV detectors.  
Indeed the GOES instrument does not observe any flux enhancement in X-rays, then this event is not classified as an X-ray flare.  Still, it presents all the EUV characteristics of a two ribbons flare with two separating ribbons and an arcade of flare loops, L$_1$, linking them (\figS{MGN171}c,f). 
Both these ribbons and flare loops are partly hidden by  other sets of 
loops, L$_2$ and L$_3$, located in front in the AIA images (\fig{MGN171}b).  
These L$_2$ and L$_3$ sets are not part of the flare which is associated to Fil$_1$ eruption. 
Rather these L$_2$ and L$_3$ sets are located over  the stable filament Fil$_2$, and they are studied in \sect{Loop}.

%#########################################################
%#########################################################

%%%%%%%%%%%%%%%%%%%%%%%%%%%%%%%%%%%%%%%%%%%%%%%%%%%
\subsection{Magnetic Configuration}
\label{sect_Magnetic}

%   {\S}{\bf --- Evolution of B in AR 11690} \\
The evolution between 2013 March 11 -- 16 of the magnetic field in the vicinity of the filaments is displayed in \fig{mag} in the local solar frame.  The main magnetic polarities are labeled in panel (e). AR 11690 is mostly in a decaying stage, with the emergence of a bipole in its negative polarity N$_1$(\fig{mag}b-d).  This induced a reorganization of its leading polarity from 12 to 14 March. 
More importantly, long-term small scale cancellations of the magnetic flux are observed at the PIL  
between N$_1$ and P$_0$ (this is best seen in the associate movie).  These cancelations are induced by the dispersion of the magnetic polarities due to sub-photospheric convection motions with a time scale of days.   
This flux cancellation is expected to build the FR which supports the filament \citep{Ballegooijen1989}. It is also at the origin of the slow evolution of the magnetic configuration which  can bring it to instability \citep{Amari2010, Aulanier2010}.  

%   {\S}{\bf --- Location of the filaments} \\
The erupting filament Fil$_1$ is located at north-west side of the AR 11690 as shown with the red contour of \ha\ observation co-aligned with the magnetogram  (\fig{mag}d).  More precisely the bottom of the \ha\ filament is following the PIL between the AR negative leading polarity N$_1$ and a dispersed large-scale positive polarity P$_0$ (the remnant of an earlier AR).  The filament extends on the positive polarity indicating that it is inclined towards the west.
{ The explanation for the westward inclination of the filament Fil$_1$ away from its underlying PIL is the asymmetry of the magnetic field strength on both sides of the PIL.  This inclination is present in MHD simulation during the equilibrium phase and even more visible during the eruption phase \citep{Aulanier2010, Torok2011, Titov2018, Titov2021}. From coronal observations, such asymmetry is also known to cause eruptions to be inclined toward the weak-field side \citep[e.g.,][]{Panasenco2013,Kay2015}. }
 Next, a  narrow stable filament, Fil$_2$, is located along the main PIL within AR 11690 (blue contour in \fig{mag}d).  The filament Fil$_2$ is following well the local PIL and it is narrower than Fil$_1$ as typically observed for filaments within ARs compare to filaments at the periphery of ARs.
 Furthermore, both filaments, Fil$_1$ and Fil$_2$, are on the same PIL located around the polarity N$_1$.  Fil$_1$ is even curving slightly inside AR 11690 in its northern part (\fig{mag}d).

%   {\S}{\bf --- Continuity of the filaments?} \\
As mentioned in Section 2.1, the shape of the observed ribbons is reverse J-shaped. Such orientation of the ribbons is an evidence of left handedness in the AR 
\citep{Demoulin1996, Williams2005, Chandra2009}. Based on this, we infer the handedness of Fil$_1$ is left-handed. 
Another method to determine the handedness of the filament is as follows.  We have identified 
the magnetic polarities at the end points of filament Fil$_1$ and found that the 
northern/southern foot-point are located in positive/negative polarities 
respectively  (\fig{mag}d).  In addition to this, the eastern/western 
side of the filaments are in negative/positive polarities respectively. Using these identifications, we have determine the handedness of the filament \citep[see figure 5 of ][]{Mackay2010} and conclude that the filament F$_1$ has a left-handed configuration. 
The same method is applied for the filament F$_2$ which is also found to be in a left-handed configuration.  However, this correspondence of handedness is not sufficient to infer that both filaments are in a { single} magnetic configuration along the same PIL.
We found during the quiet phase, before 12:40 UT on March 16, it is difficult to find a separation 
between the two filaments. 
{ However, the eruption of Fil$_1$ and not of Fil$_2$ is a clue supporting separated magnetic field configuration, with the limitation that filament plasma outlines only a small fraction of the magnetic configuration.}
 Then, we will first consider below two nearly parallel filaments, before coming back  in \sect{Physical_Loop} to the details of this northern part of the PIL where they are nearby.   

%   {\S}{\bf --- Interaction with AR 11691} \\
Finally, another AR, NOAA 11691, is developing southward to AR 11690 and with the same global bipole orientation (P$_3$-N$_2$, \fig{mag}).  
Since AR 11691 is nearby AR 11690 and the erupting magnetic structure is strongly expanding southward, the coronal field of AR 11691 partly reconnects with the erupting magnetic field (see \fig{MGN171}e, related movie, and \sect{Overview}).

\begin{figure*}[t!]    % Figure 4
\centering
\includegraphics[width=0.9\textwidth]{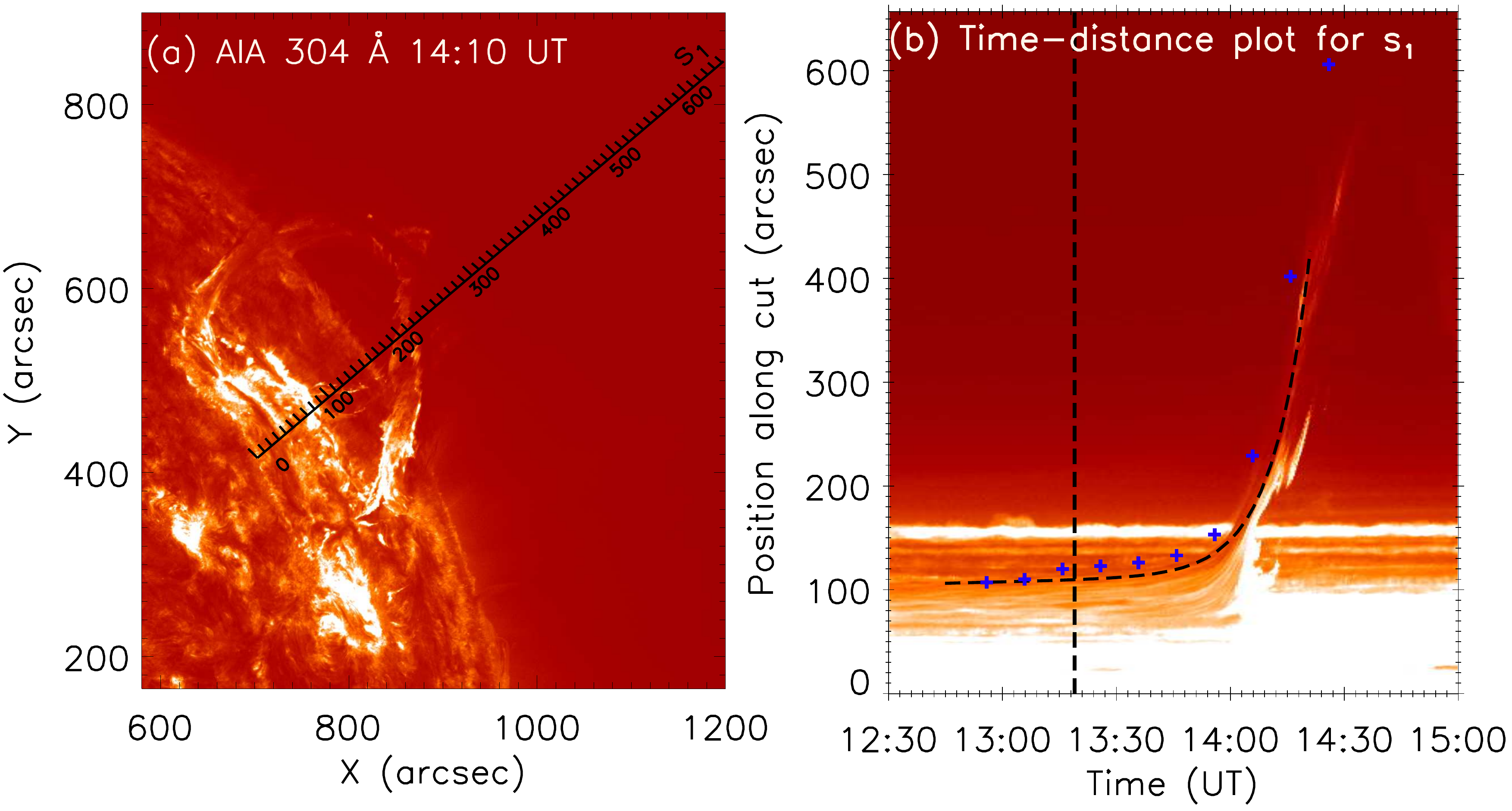}
\caption{
(a) AIA 304 \AA\ image with the location of slice S$_1$ used for the analysis of the 
filament Fil$_1$ height-time evolution.  
(b) Height-time plot along the slice S$_1$. The height is corrected from solar rotation.
The vertical dashed line indicates the approximate onset time of the filament eruption. 
The black dashed line is the fit of \eq{h(t)} to the filament apex as traced with the 
AIA 304 \AA\ data along slice S$_1$. 
The blue crosses represent the apex height of the filament leading edge derived by 
triangulation with the data of SDO/AIA and STEREO~A/EUVI (see \figs{evolution}{stereo}). 
The height is set at the same position for the first blue cross. During the eruption these 
heights are slightly above the AIA 304 \AA\ apex and its fit with \eq{h(t)} 
due to a projection effect.  
}
\label{fig_ht}
\end{figure*}
%#########################################################

%%%%%%%%%%%%%%%%%%%%%%%%%%%%%%%%%%%%%%%%%%%%%%%%%%%
\subsection{Eruption Kinematics}
\label{sect_Kinematics}

%   {\S}{\bf --- Define slice S$_1$ for eruption kinematics} \\
We explore the eruption kinematics using AIA data sets.  A global and qualitative view of the eruption is given by the three movies attached to \fig{evolution}.  A quantitative analysis is done with the slice S$_1$ shown in \fig{ht}a.  
We define the location of slice S$_1$ in AIA images to follow the filament leading edge.  The measured height is corrected from solar rotation.
While the filament eruption and its consequences are seen the best with 171 \AA\ data, the early part of the 
eruption is partly masked with the sets L$_2$ and L$_3$ of coronal loops present in front of the 
erupting filament (middle left of \fig{MGN171} a, b).  These loops are not emitting in 304 \AA , then  
we privilege this filter to study the pre-eruption phase, as well as the whole filament 
eruption, while we also apply the same method to the other filters for completeness.

%   {\S}{\bf --- Fitting procedure} \\ 
The kinematics of the filament eruption is an important clue for understanding 
the triggering mechanism of the eruption. Therefore, in order to analyze the kinematics of the filament eruption, we visually select points along the bright tracks of the time-distance plots as shown in \fig{ht}b.  \citet{Cheng2020} have analyzed 12 eruptions
and test several functions to fit the data. They found that the combination of a linear and exponential time dependences provide the best description of the full data set, so a lowest $\chi^2$ for most eruptions, compare to previously suggested functions (like a power law of time).  The most general equation analyzed by \citet{Cheng2020} is
   \begin{equation}  \label{eq_h(t)}
    h(t) =  a\, e^{b (t-t_0)} +c\,t + d  \,.
   \end{equation}
Here, we note that $t_0$ is a redundant parameter as the first term can be rewritten 
as $a\, e^{-b t_0} e^{b t} = a' e^{b t}$ so that all the combinations of the three 
parameters $a,b,t_0$, which provides the same $a'$ value, define exactly the same 
function.  Said differently, $b$ is defined by the temporal behavior of the data 
while changing $a$  in \eq{h(t)} could be exactly compensated by changing $t_0$.   Then, we fix $t_0$ to a given value, 12:45 UT { as \citet{Cheng2020} fixed it to their first data point. } 
The coefficients  $a, b, c$, and $d$ are determined by minimizing the reduced $\chi^2$, $\chi^2_{r}$, between the data and \eq{h(t)}.
The fitting is done using the mpfit routine available in the solar software (SSWIDL). 
The goodness of the fitting is given by the reduced $\chi^2_{r}$ value obtained. 
The $\chi^2$ is defined by $\chi^2$ = $\sum_{i=1}^{N}$ ${[h_i(t) - H_i(t)]^2}$, where h$_i$ and H$_i$ are the fitting and measured heights, respectively and ‘t’ is the time.  
This formula is  slightly different than the one used by \citet{Cheng2020} since we are not including the error for each measured height (the minimum found assumed the same error for each measurement). Finally we calculated the reduced  $\chi^2_{r}$ value as $\frac{\chi^2}{DOF}$, where DOF is the number of the degree of freedom.  We find the minimum $\sqrt{\chi^2_{r}}= 1.4\arcsec $. This is only slightly above 
two pixels of AIA ($1.2\arcsec $), indicating that \eq{h(t)} provides a close representation of the data.
 
%   {\S}{\bf --- Kinematics of the eruption} \\ 
During the earlier times of \fig{ht}b, a slow linear increase of the filament height is present, defined as the slow-rise phase by \citet{Cheng2020}.  The fit of \eq{h(t)} to the data provides $c \approx 2$ \kms ~ in all AIA channels.  This slow-rise phase is typically associated with the presence of weak brightenings. They are interpreted as the consequence of tether-cutting reconnection which allow the slow upward motion.
Later on, during the acceleration phase, the speed of the eruption increases exponentially.
The maximum speed of the eruption was computed by derivating the height-time fit of the data. The calculated maximum speed reached  up to 
300 \kms\ within the AIA field of view. This increase occurs on the time scale  1/b $\approx$ 9.6 min. 
This exponential behavior is characteristic of the linear development of an instability.
 
%   {\S}{\bf --- Onset time} \\
The beginning of the eruption is difficult to define precisely as the filament top is smoothly changing from a linear 
to an exponential rise.  Also, there is no characteristics time defined by \eq{h(t)} as $t_0$ is an ill defined 
parameter (see above), and then cannot be associated to an onset time.  Indeed, for both a linear and exponential behaviors there is no specific time which can be referred to (said differently both function can be shifted in time while keeping the same form).  Then, an extra information needs to be added to define the beginning of the eruption from \eq{h(t)}.  The instability starts at least before the exponential rise is large enough to be detected.  This is best quantified with the velocity, $a\,b\, e^{b t}$ becoming larger than a threshold value $v_t$.  Then, the eruption start time is defined as:

  \begin{equation}  \label{eq_t_start}
    t_{\rm start} =  \frac{1}{b} \ln \left(\frac{v_t}{a\, b}\right)  \,.
   \end{equation}

Taking $v_t = c$ sets the exponential velocity equal to the linear velocity, so the growth of the instability well visible above the linear rise.  This occurs at $\approx$ 13:30 UT.  The eruption is expected to starts even earlier on.  The exponential is well detectable over the fluctuations of the slow-rise phase when it is larger than three times the standard deviation, which is found to be $\approx  0.62$ \kms .   Including this velocity as $v_t$ in \eq{t_start} defines the eruption start at 13:19 UT, so 11 min before.  This last time is expected to be closer to the real eruption start, which is expected to be even earlier on when the exponential was too small to be detectable.  To be conservative, we set the starting time at 13:19 UT. 

%#########################################################
\begin{figure*}[t]    % Figure 5
\centering
\includegraphics[width=0.9\textwidth]{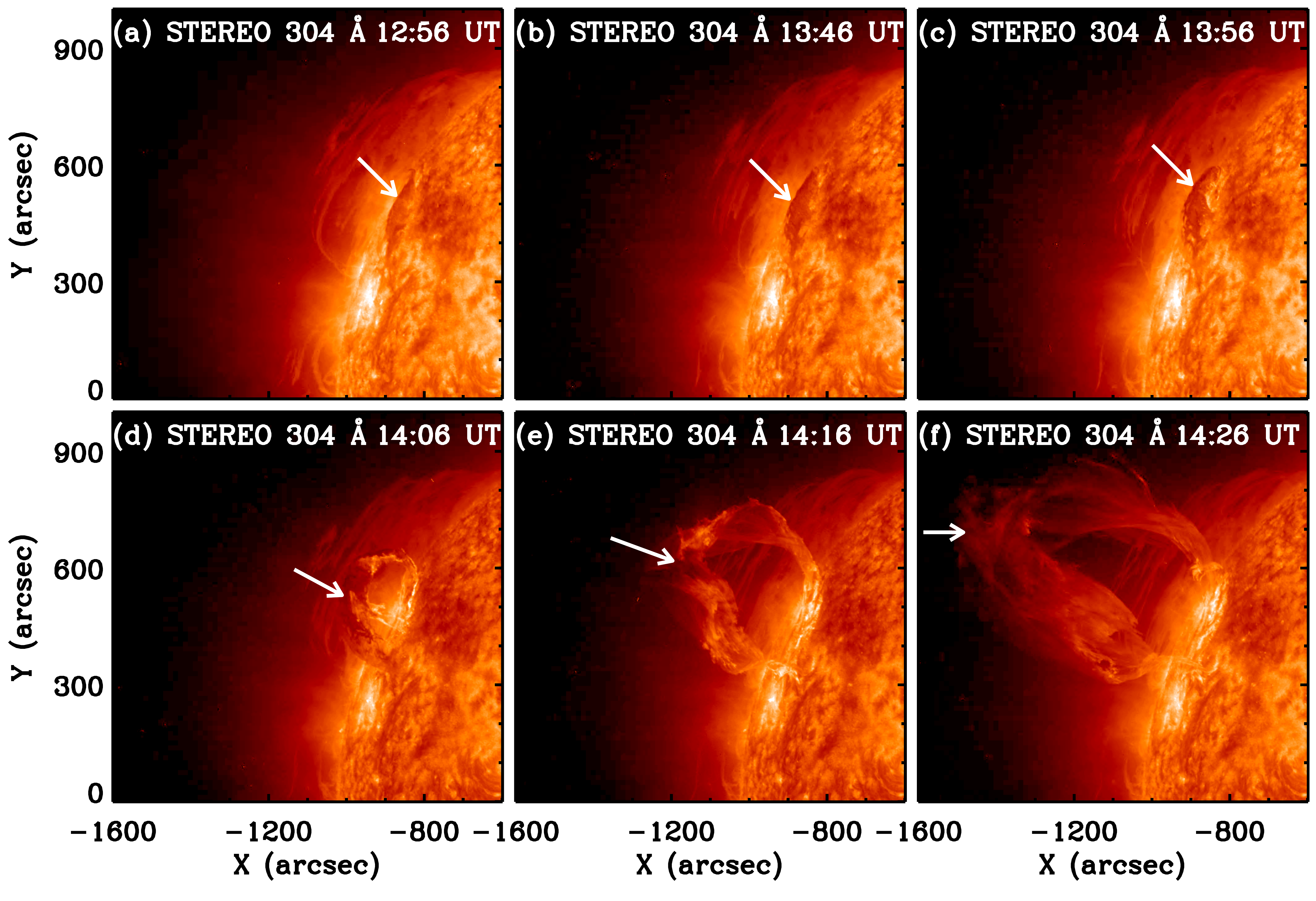}
\caption{Evolution of the filament observed by STEREO~A~/EUVI in 304 \AA\ at the eastern solar limb.  
The view point is on the other side of the filament as observed with AIA (\fig{ht}a) with a longitude 
difference of 130 $\degree$ between the spacecraft. The filament/prominence leading edge apex is 
indicated by arrows.} 
\label{fig_stereo}
\end{figure*}
%#########################################################

%   {\S}{\bf --- STEREO~A observations} \\
The eruption was also well observed by the STEREO~A spacecraft with a different viewing angle in longitude. 
The angle between the STEREO~A and the SDO on 2013 March 13 was 130$\degree$ and 
for STEREO~A the filament was located close to the eastern limb. 
The evolution of the eruption in 304 \AA\ is shown in \fig{stereo}.  The arrow points to the leading edge apex 
of the eruption. This leading edge is a smooth bent curve at the beginning of the eruption as with 
SDO observations (\figs{evolution}{MGN171}). The filament feet appearance below the filament are partly different than the ones observed by AIA because STEREO~A observes the other side of the filament.   As the eruption proceeds, the filament becomes bright both because of plasma heating, as described above for AIA observations, and because the plasma is observed over the limb with a weaker emitting background, so as a prominence. 
Next, a large quiescent prominence is present in the background at the eastern limb (\fig{stereo}). This corresponds to an extended filament, about one solar radius 
long, located in between the diffuse eastern extension of the positive polarity of 
AR 11690 (well outside the field of view of \fig{mag}) and another large scale and 
diffuse negative polarity located further to the East.  No significant consequence 
of the eruption of Fil$_1$ is observed on this large quiescent filament (\fig{stereo}).

%#########################################################
\begin{figure*}[t!]    % Figure 6
\centering
\includegraphics[width=0.9\textwidth]{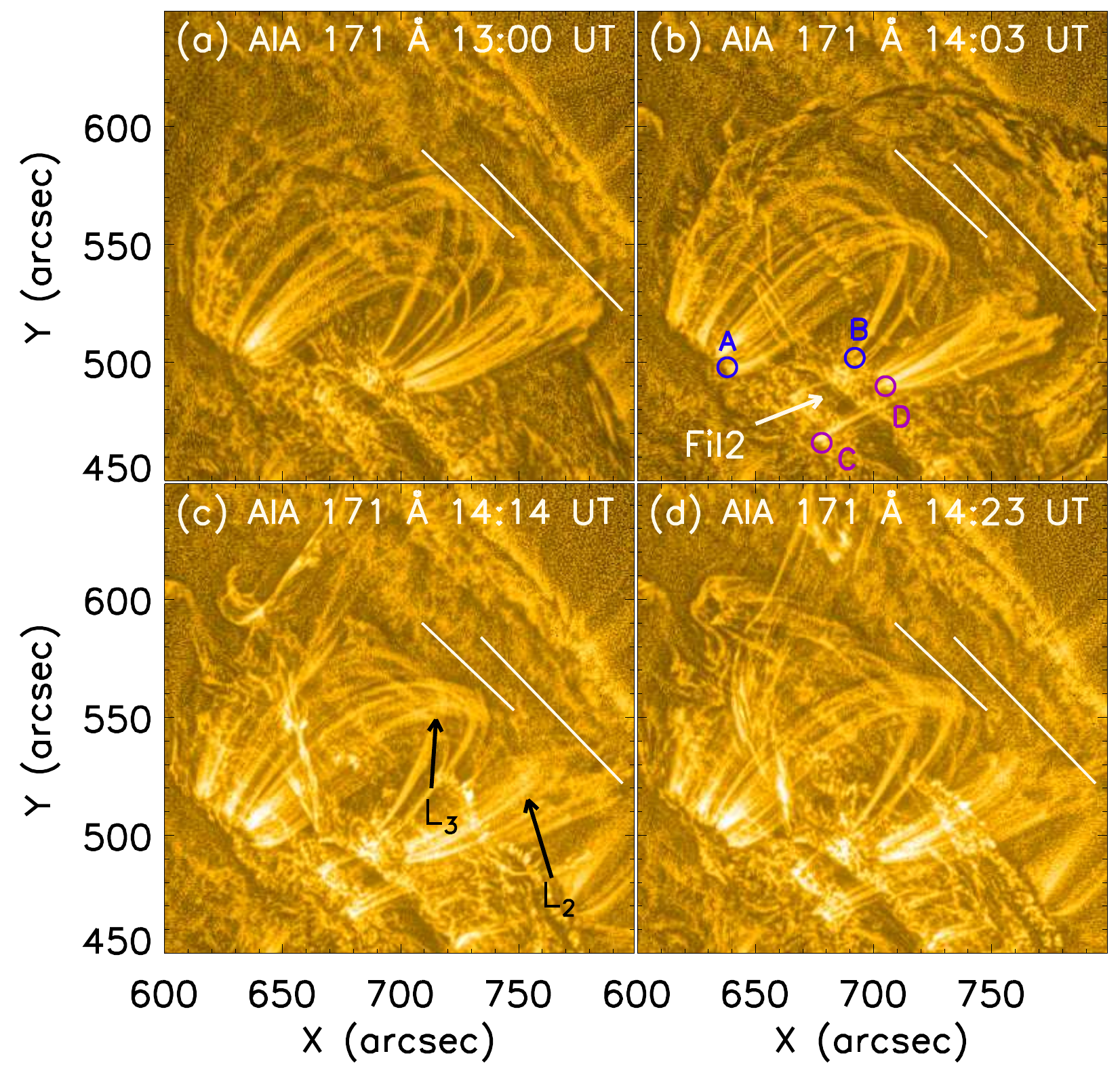}
\caption{Evolution of the two loop sets located in the neighborhood of the filament Fil$_1$ eruption 
and above Fil$_2$ (seen in the top part of panel (b)). 
The data are from AIA 171 \AA\ processed with the MGN technic.  The white lines are at fixed positions in the solar frame to better visualize the loop evolution.  After the filament eruption onset at $\approx$  13:19 
UT, the loops first contract (panel b), then they expand backward towards their initial locations (panels c and d). 
The two loop systems are labeled with L$_3$ and L$_2$ in panel (c). The selected loop  footpoints are marked with A, B, and C, D, respectively, in panel (b), and they are reported in \fig{mag}f}.

\label{fig_loops}
\end{figure*}
%#########################################################

%   {\S}{\bf --- SDO-STEREO~A triangulation } \\
 By combining the data of AIA 304 \AA\ and STEREO~A 304 \AA\ data, these stereoscopic observations allow to derive the true height of the filament.  The location pointed by the arrow in \fig{stereo} is taken for the computation of the filament top height using the trigonometric triangulation method ``$ssc\_measure$'' available in solar soft. 
The points selected for the height-time plot in triangulation method are along the direction of slice S$_1$ for AIA observations. This provides
us an opportunity to compare the calculated speeds derived by the time-distance analysis and the stereoscopic analysis. 
For the comparison of the height derived from the triangulation method and from the AIA time-distance 
analysis, the common height reference is set before the eruption at the top of the filament.
The calculated heights are overplotted on the time-distance plot of \fig{ht}b with the blue `+' symbol. Using these points, we compute a mean speed of $\approx 180$ \kms.
  This derived speed from triangulation methods is about 20 $\%$ faster than the speed derived from the time-distance analysis of the AIA data. 
This difference is due to projection effects as the velocity measured with AIA 
does not include the velocity component out of the plane of sky.

%   {\S}{\bf --- CME } \\
  Finally, the filament eruption produced a CME observed by {\emph SOHO/LASCO} in the north-west direction. 
In LASCO C2 field of view the CME appear $\approx$ 14:48 UT at a height of 4 \Rsun\  and it is visible in LASCO C3 field of view up to $\approx$ 25 \Rsun\ at 19:54 UT. The CME was a partial halo having a projected width of 323$^{\circ}$. The measured mean speed within C2 and C3 fields of view is $\approx$ 790 \kms\ (see \url {https://cdaw.gsfc.nasa.gov/CME\_list}),  then the ejection was further accelerated compare to the above measurements in the low corona.

%%%%%%%%%%%%%%%%%%%%%%%%%%%%%%%%%%%%%%%%%%%%%%%%%%%
\subsection{Loops Contraction and Expansion} 
\label{sect_Loop}

%{\S}{\bf --- Global description of the present observations} \\
In this section, we analyse the EUV loop contraction and expansion related to the eruption of the filament Fil$_1$ on 2013 March 16.  
We will not study the shrinkage of flare loops L$_1$ as this phenomena was already well studied (see Section \ref{sect_Introduction}) but rather the evolution of coronal loops which are not involved in the flare reconnection.
This phenomena is observed with 171 and 193 \AA\ filters in two set of loops (\fig{loops}).   
We name these two loop systems as L$_2$ and L$_3$, respectively, as indicated in Figures~\ref{fig_MGN171}b 
and \ref{fig_loops}c.  Both loop systems are rooted in $P_1$ (positive) and $N_1$ (negative) 
polarities (\fig{mag}e,f) of AR 11690. The location of STEREO A compare to SDO is suited for a 
triangulation, as done above for Fil$_1$, however this is not possible for the loops of  L$_2$ and L$_3$ as they are hidden behind the filament Fil$_1$ (\fig{stereo}).
 
%	{\S}{\bf --- Description of L$_3$ and L$_2$} \\
  The eastern set, L$_3$, is better observed with loops seen from one side, nearly face on, and 
they are well defined over their full length. These loops are located above the location 
where filament Fil$_1$ and Fil$_2$ are nearly joining  (\fig{loops}b,c).
 They end on both sides of these filaments  (\fig{mag}d,f). 
The geometry of the western set L$_2$ is more difficult to define from AIA observations since the 
direction of observation is nearly along the loops (side on).  Moreover, the part closer to the 
observer is faint, and the emitted light is mixed with the one of the background coming from 
the stable filament Fil$_2$ and its surrounding brightenings.
{Still, the bottom part of these legs of the loop system L$_2$ can be seen on the front side of the filament Fil$_2$.}
Then, the loops L$_2$ are rooted on both sides of filament 
Fil$_2$, as the loop system L$_3$  (\fig{mag}d,f).   
The corresponding co-aligned photospheric magnetograms, \eg\ \fig{MGN171}d, on March 16 confirm 
that both sets of loops are rooted in the magnetic polarities surrounding the stable filament 
and the end of the erupting one.  With a space filling coronal magnetic field (low plasma $\beta$ conditions), L$_2$ and L$_3$ belong to the same magnetic arcade passing over Fil$_2$ and extending all along polarities $P_1$  and $N_1$.  
The heating is probably not large enough in the arcade middle to create dense enough coronal plasma, then there is a gap of coronal emission and the appearance of two separate sets of loops.  
 We conclude that these two sets of loops belong to the magnetic arcade which is overlaying the stable 
filament Fil$_2$ outlined with a blue contour in \fig{mag}d and end above the northern end part of filament Fil$_1$. 

%   {\S}{\bf --- Loops evolution. \fig{loops}} \\
The evolution of these loops during the filament eruption is shown in \fig{loops} at four times with AIA 171 \AA\ filter.  We have drawn two straight lines at the top of the loops at 13:00 UT (before the eruption onset).  These lines are repeated in the next panels with fixed coordinates in the local reference frame, so by taking into account the solar rotation. Then, these lines allow to better visualize the loop evolution.
 We also refer to the attached movies at AIA 171 and 193 \AA\ wavelengths to view the full evolution of these loops.  
These movies show that all these loops are contracting in phase.   The contraction is maximal about 43 min after the filament eruption onset (\fig{loops}b). Later on their motions reverse and they expand approximately back to their original positions.  No further oscillation is observed.

%#########################################################
\begin{figure*}    % Figure 7
\centering
\includegraphics[width=0.8\textwidth]{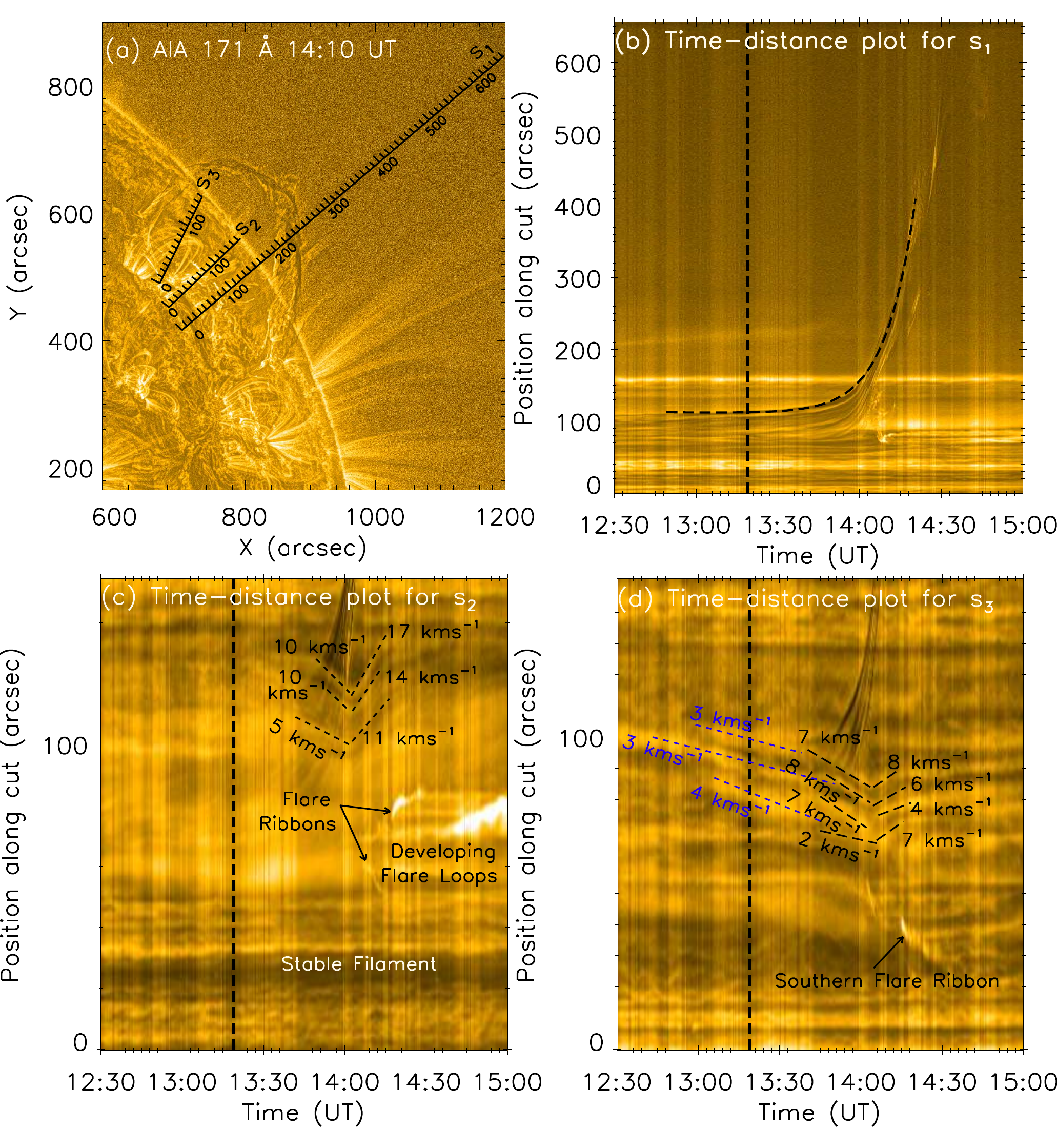}
\caption{
  (a) AIA 171 \AA\ image processed with the MGN technic.  The location of three slices are shown. Slice S$_1$ 
is the same as in \fig{ht}a.  It monitors the filament height versus time.  Slices S$_2$ and S$_3$ are 
monitoring the loop contraction/expansion along the loop systems L$_2$ and L$_3$, respectively (see \fig{loops}). 
  (b) Height-time plot for the filament eruption with the fit of \eq{h(t)} to the AIA 171 data of the 
filament leading edge added with a dashed line. 
  (c, d) Height-time plots along slices S$_2$ and S$_3$ showing the contraction and expansion of the EUV loops during the filament eruption. The vertical dashed line indicates the approximate onset time of the eruption. 
The short dashed segments outline the loop contraction/expansion of the loops.  The derived mean speeds are added close to each 
segment.  In panel (c) the trace of the flare ribbons and of the developing flare 
loops are indicated. In both panels (c, d) the erupting filament is seen when it 
emerged from the occulting coronal loops L$_2$ and L$_3$. 
 } 
\label{fig_slice_171}
\end{figure*}
%#########################################################

%#########################################################
\begin{figure*}   % Figure 8
\centering
\includegraphics[width=0.8\textwidth]{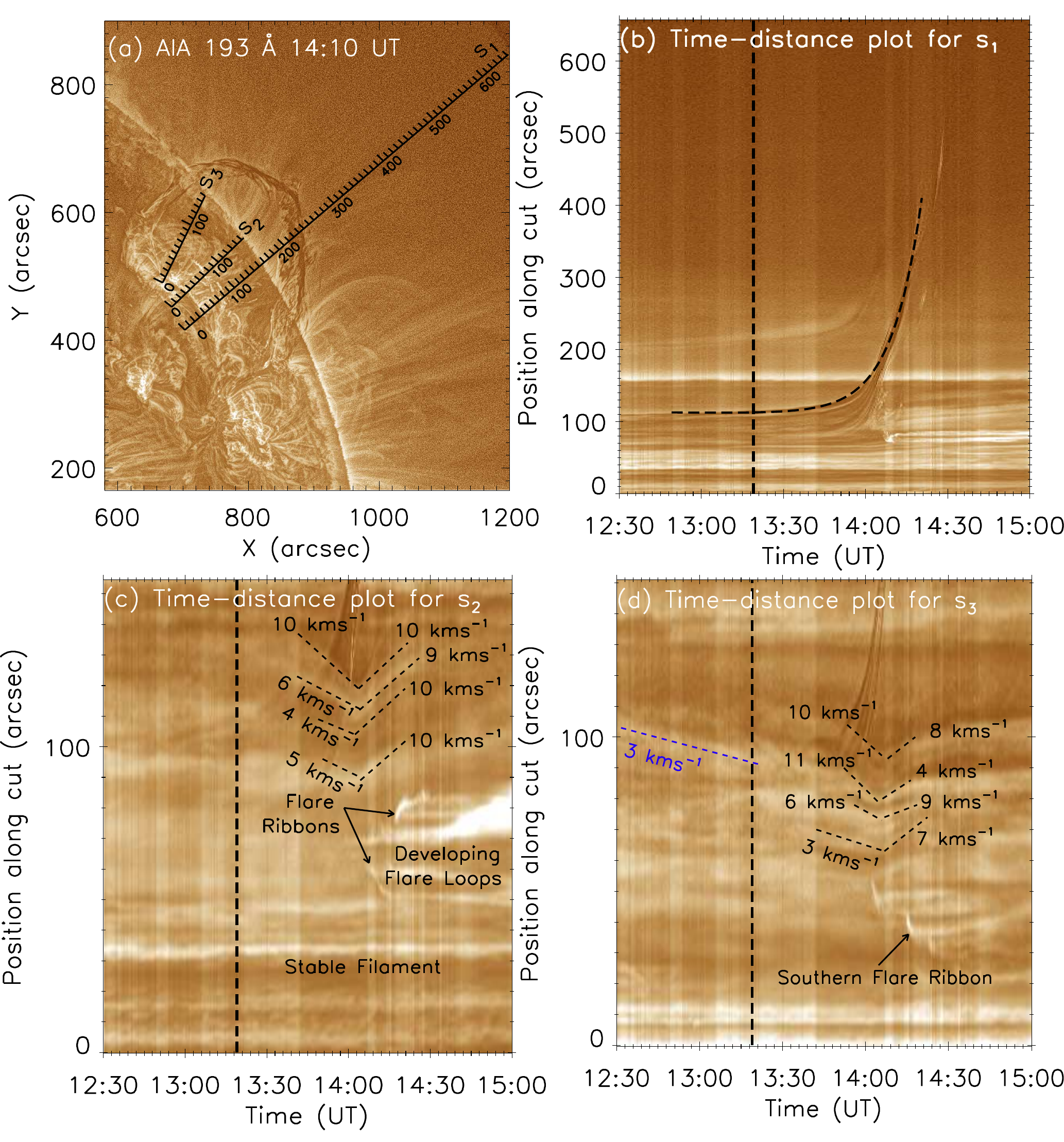}
\caption{Data of AIA 193 \AA\ processed with the MGN technic and with the same format as \fig{slice_171}.   
  (a) AIA 193 \AA\ image with the same slices (S$_1$, S$_2$ and S$_3$) than in \fig{slice_171}a.
  (b) Height-time plot for the filament eruption with the fit of \eq{h(t)} to the AIA 193 data of 
the filament leading edge added with a dashed line. The height has a constant projection factor on the plane of sky.
  (c, d) Height-time plots showing the contraction and expansion of the EUV loops after the filament eruption along slices S$_2$ and S$_3$.  The vertical dashed line indicate the onset time of eruption. 
The typical contraction/expansion speeds of the loops are added, as well as the flare traces.} 
\label{fig_slice_193}
\end{figure*}
%#########################################################

%   {\S}{\bf --- Slice definition} \\
For a quantitative analysis of these two sets of loops, we show the results obtained along two slices selected to cross the loop tops nearly orthogonally, so along the loop motion.  We name these slices S$_2$ and S$_3$.  They cross the loop sets L$_2$ and L$_3$, respectively. The results of these space-time analysis is presented in \figs{slice_171}{slice_193} for the AIA 171 and 193 \AA\ data, respectively.  We corrected the height projection from the solar rotation. The results with slice S$_1$, defined in \fig{ht}a, are also included in the panels (b).  They both confirm the upward motion and velocities of the filament Fil$_1$ obtained with the 304 \AA\ filter.  In contrast, the stable filament Fil$_2$ is present at a nearly constant position in the lower part of panels (c) of \figs{slice_171}{slice_193}. The upward little drift of Fil$_2$ is not due to the solar rotation as it is removed.
We interpret this drift as follows. This stable filament is in a decaying AR, and cancelation occurs at the PIL, further building up the FR configuration. 
So the FR is slowly moving up towards a nearby equilibrium. 
Such slow evolution has been analyzed before for several observed events \citep[\eg\ ][]{Byrne2014, Gosain2016, Chandra2017}.

%   {\S}{\bf --- slice S$_2$ at 171 \AA\ } \\
The loops L$_2$ crossed by slice S$_2$ are initially stable up to $\approx$ 13:40 UT when 
they start contracting (\fig{slice_171}c).  At that time, the apex of filament Fil$_1$ was 
already moving upward  at a velocity $\approx 6$ \kms\  (\fig{slice_171}b).  
Across slice S$_2$,  the erupting filament Fil$_1$ is partly masked by the foreground set of loops L$_2$ present along the line of sight (Figures \ref{fig_loops}, \ref{fig_slice_171}c).  Fil$_1$ is observed in absorption before and after the crossing of the contracting loops.  
The contraction of all the loops starts almost simultaneously  $\approx$ 21 min after the 
 filament eruption onset.  This is in the range 4 to 40 min of the time difference between the onsets of filament eruption and loop contraction reported before  \citep{Shen2014, Dudik2016, Dudik2017, Wang2018, Devi2021}. 
%\pc{from the paragraph suppressed in Section 3.2}
The speed of the contraction varies from 5 to 10 \kms.  This contraction continues for a duration of $\approx$ 22 min.  At $\approx$ 14:02 UT, these loops change suddenly from contraction to expansion without any significant phase difference between them.  The speed of the expansion is in the range 11 to 17 \kms . This expansion speed is higher by a factor 2 than the contraction speed.  The expansion continues at least for 15 min, and afterwards these loops L$_2$ are difficult to detect with the 171 \AA\ filter.

A similar loop evolution is present with AIA 193 \AA\ (\fig{slice_193}).
The speeds of the contracting and expanding loops are comparable to the one observed in AIA 171 \AA . The small difference in velocities could have several origins: the difference in the emitting 
plasma, the diffuse appearance of the loops, the short and partly different duration of the 
contraction and expansion phases (as could be observed when enough emitting plasma was present). 
Finally, one more loop, with similar time evolution than others, is detected in 193 \AA\ (the lower one in \fig{slice_193}c).

%   {\S}{\bf --- slice S$_3$} \\
The slice S$_3$ analyses a different set of loops, L$_3$, which are located closer to the northern leg of the erupting filament Fil$_1$ (\fig{slice_171}a).  This loop system has a similar behavior as the loop system L$_2$ except that it shows also a contraction before the filament eruption onset. We refer this contraction as `earlier loop contraction'. This earlier loop contraction can be clearly seen in the panels (d) of \figs{slice_171}{slice_193} respectively (see also the accompanying movie). This contraction starts between $\approx$ 12:30 and 13:00 UT and the computed speed for the earlier contraction is about 3--4 km s$^{-1}$.
Next, about 21 min after the eruption onset, the loops in S$_3$ contract  as the ones of slice S$_2$.  The contraction speed in the case of 171 \AA\ varies from 2 to 8 \kms\ which is comparable to the speeds deduced from the 193 \AA\ filter (in the range 3 to 10 \kms).  
The contraction time for both wavelengths is about 22 min.  As the loops of S$_2$, the contraction changes rapidly to an expansion at $\approx$ 14:02 UT.   The expansion speed varies from 4 to 8 \kms\ in case of 171 \AA\ and from 4 to 9 \kms\ in the case of 193 \AA\ so comparable to the contraction speed.  For both contraction and expansion, there is a global tendency of an increasing velocity with height, a tendency which is also present for slice S$_2$.  Next, at the difference with loops L$_2$, loops L$_3$ could be followed much longer in time.  About 20 min after motion reversal, this expansion slows down and it is over by $\approx$ 14:27 UT (panels (d) of \figs{slice_171}{slice_193}). No significant evolution is present later on. 

%   {\S}{\bf --- Flare consequences} \\
The slice S$_2$ shows also the consequences of the flare reconnection (panels (c) of \figs{slice_171}{slice_193}). 
The flare ribbons start to significantly brighten and to separate from each others in slice S$_2$ at $\approx$ 
14:18 UT, while they are detected starting earlier, at $\approx$ 13:59 UT, close to slice S$_3$ which is coherent with an eruption starting at the end of Fil$_1$.  The flare loops are observed 
later on, after $\approx$ 14:34 UT in S$_2$.
The drift of position observed in panels (c) of \figs{slice_171}{slice_193} is interpreted as the formation of higher flare loops as magnetic reconnection proceeds.
In slice S$_3$, after 14:00 UT and for an abscissa $< 50$ \arcsec there is a brightening shift to lower S$_3$ abscissa.
We have examined this shift in detail by analyzing the movies and corresponding images.  This shift corresponds to the motion of the southern flare ribbon.

\section{Physical Interpretation}
\label{sect_Physical}
%%%%%%%%%%%%%%%%%%%%%%%%%%%%%%%%%%%%%%%%%%%%%%%%%%%%%%%%%%%%%%%%%%%%%%%%%%%%%%%%%%%%%%

%%%%%%%%%%%%%%%%%%%%%%%%%%%%%%%%%%%%%%%%%%%%%%%%%%%
\subsection{Erupting Magnetic Field Configuration}
\label{sect_Physical_Magnetic}

%	{\S}{\bf --- How Fil$_2$ is forming ?} \\
   The evolution of the radial magnetic field component is typical on an AR in decay (\fig{mag} and associated movie).  Magnetic flux is progressively dispersed by super-granules convective cells.  This implies that both magnetic polarities of AR 11690 are growing in extension.  This induced magnetic field cancelations around the photospheric level along the internal PIL of the AR.   This evolution is generically expected to build a FR \citep{Ballegooijen1989, Aulanier2010,Green2011} where dense plasma could be caught \citep[\eg\ ][]{Aulanier1998}, forming filament Fil$_2$.
   
%	{\S}{\bf --- How Fil$_1$ is forming ?} \\   
Photospheric magnetic cancelation is also triggered by the field dispersion at the AR boundary, especially at the PIL between polarities N$_1$ and P$_0$.  
 Our observations of cancellation of magnetic flux at the photospheric level below the filaments are consistent with the study of the filament formation of \citet{Wang2007}. Two of their studied filaments were formed at the periphery of ARs, like Fil$_1$, while the third one was formed inside a decaying AR like Fil$_2$. 
They concluded that the filaments are formed due to the same process of magnetic flux cancellation at photospheric inversion line.
Here, the studied filaments, Fil$_1$ and Fil$_2$, are around the same magnetic polarity N$_1$. Before the eruption it is difficult to set the limit between Fil$_1$ and Fil$_2$ but eruption of only Fil$_1$ { favours that they have two separated magnetic configurations.}
%	{\S}{\bf --- Consequences of the emerging bipole} \\

  The main add to the above magnetic diffusion scenario is the emergence of a magnetic bipole in the 
leading negative polarity of AR 11690 (\fig{mag}b-e and associated movie).  This bipole has almost 
the reverse orientation than the main bipole forming AR 11690 (P$_1$,N$_1$).  The emerging positive 
polarity mainly cancels with the outer (westward) part of polarity N$_1$.  This reconnection transfers 
the negative footpoint from N$_1$ to the emerging negative polarity.   The coronal implication of 
this reconnection depends on the initial coronal connectivity of N$_1$.  The part closer to the 
internal PIL of AR 11690 is expected to connect to P$_1$.  In this case the reconnection brings 
the negative footpoint of the reconnected fields connecting P$_1$ closer to the internal PIL. 
This  decreases the curvature radius of field lines, then strengthen the stabilizing 
 magnetic tension force of the  magnetic configuration supporting filament Fil$_2$.  However, the external part of N$_1$ is expected to connect P$_0$ (\eg\ as it does with a potential field).  In this case the reconnection brings the negative footpoint of the reconnected fields connecting P$_0$ away from the external PIL.
This weaken the stabilizing  tension force of the filament Fil$_1$, then it is an 
ingredient to bring the magnetic configuration of Fil$_1$ to eruption.  
Still, this reconnection was not sufficient since the positive polarity of the 
emergence canceled and disappeared on day before Fil$_1$ eruption.
{ These results are in agreement with the study of \citet{Chen2000}.}

%	{\S}{\bf --- Link to the standard model} \\
  The 3D standard model of eruptive flares is typically developed in a bipolar field modeling an AR with the build up, then eruption, of a FR formed above the internal PIL \citep[\eg\ ][]{Aulanier2010,Janvier2015}.  
In the case of an external PIL, located at periphery an AR, the same model could be qualitatively be 
applied if a first coronal reconnection forms a sheared arcade over this PIL, as shown by \citet{Torok2018}.  
The following evolution is mainly driven by the diffusion of magnetic polarities, as for the case of an internal PIL.  The main difference at this stage is an expected slower process as the spatial region involved around the external PIL is large while convective cells have typically the same speed.   

%	{\S}{\bf --- Instability} \\
  When the magnetic configuration reaches an instability (as evidenced by the exponential growth of the filament height), the filament trace an erupting FR structure  well observed both with SDO/AIA and STEREO A/EUVI 304 \AA\ filters (\figs{ht}{stereo} and related movies).  Reconnection behind the erupting FR leads to the formation of a flare loop arcade ending in two J-shaped flare ribbons observed in EUV wavelengths. They separate as a function of time as excepted in the standard 3D eruption model. 
   Then, we conclude that, while located at the periphery of an AR, the eruption of Fil$_1$ has all the characteristics expected with the 3D standard model of eruptive flares build for eruptions located in the core of ARs.  The main differences are slower processes both for the formation and the ejection of the FR (due to a weaker magnetic field, so weaker forces).

\subsection{Contracting and Expanding Loops}
\label{sect_Physical_Loop}

{ The loop system L$_3$ (panel (d)   of \figs{slice_171}{slice_193})  shows the earlier contraction as mentioned in Section \ref{sect_Loop}.  The data presented here provide no clue about this contraction. We observe small brightenings at the base of loop system L$_3$ before the eruption and a short description of our analysis is presented in appendix \ref{sect_Earlier_Contraction} since this could be interesting for further studies on the subject.}
 In present observations the sets of contracting loops, L$_2$ and L$_3$, mostly recover their 
original heights they had when the fast contraction started
(\fig{loops}, panels (c) and (d) 
of \figs{slice_171}{slice_193}).  This evolution is different than for loops located in the elbow 
of AR sigmoids as, the expansion following contraction is frequently not able to recover the 
pre-eruption location \citep[\eg\ ][]{Liu2012,Simoes2013,Shen2014,Wang2018} while in some cases 
it does \citep[\eg\ ][]{Dudik2017}.  
 This behaviour is expected for loops at a remote filament channel where no energy release happens.
A possibility is that the eruption generates a coronal wave 
which first pushes the loops L$_2$ and L$_3$ downward, then the loops recover their initial positions after the coronal wave passage. Indeed, there have been previously studied cases where loops oscillate during the crossing of a coronal wave \citep[\eg\ ][]{Ballai2007,Guo2015,Fulara2019}.
In these studies, the oscillations have at least one cycle. In the present case, a weak coronal wave is associated with the eruption as visible at \url {http://suntoday.lmsal.com/sdomedia/SunInTime/2013/03/16/AIAtriratio-211-193-171-2013-03-16T1200.mov.mp4}. Due to the weak nature of coronal wave, it is very difficult to estimate if it interacted with the loops and at which time. 
The main observational constraint is that the loops have only half period oscillation with a triangular shaped amplitude.   This is far from the behavior expected from an unforced oscillator, as summarized \eg\ in Figures 2 and 4 of \citet{Russell2015}. The observed triangular shape in \figs{slice_171}{slice_193} is also well different from oscillations observed in other events \citep[\eg\ ][]{Gosain2012, Liu2012, Simoes2013}.  Then, an excitation of the loops by a coronal wave is doubtful.  

{
An additional possibility for the origin of the loop contraction and expansion is a perturbation by the lateral expansion of the erupting flux of the filament.  It is known that some eruptions show a so-called overexpansion, a stronger growth of the minor radius (thickness) of the erupting flux compared to the major radius (height) \citep{Patsourakos2010a, Patsourakos2010b, Veronig2018}.
In the case of \citet{Veronig2018}, the lateral overexpansion of the CME bubble first pushes neighboring loops to the side, which leads to very clear oscillations. Subsequently, the bubble acquires a mushroom-like shape, which presses some loops to the north of the eruption downward, followed by a recovery of the loops to the original height. Such behavior was so far seen preferentially  in fast or impulsive eruptions. However, a similar dynamics in slower eruptions, like the case here, might nevertheless be possible. 
Since the perturbation is temporary, a return of the loops to their original position would result naturally.

This event shows, as expected by filament models, that the erupting flux is much thicker and extends much higher than the filament in the hotter channel 211 \AA . A diffuse front is seen to rise synchronously with the filament at nearly twice the height (panel (j) of Figure 1). The synchronous motion suggests that the diffuse front is part of the erupting flux, probably at its top edge. This implies that most of the erupting flux is already higher than the loops L$_2$ and L$_3$ when their contraction begins. The lateral expansion of the flux at this stage should perturb the flux above these loops. When, as the overlying flux is pushed sideward by the CME bubble, the loops may experience a sideward and downward push.
Another possibility of this moving diffuse front at the top of the
 erupting FR can be formed by the erupting FR,  which collects, compress and
heats surrounding coronal plasma (and magnetic field too) in front of the
 FR.  So it could not be a part of the erupting flux but coronal plasma collected on the way. It is likely the beginning of a sheath formation, a structure well observed in-situ in front of magnetic clouds and more generally ejecta far away from the Sun.
}

%	{\S}{\bf --- Other mechanism?} \\
We next analyse how other proposed mechanisms  may explain these observations.  
The implosion conjecture is difficult to test since it was so far not quantitatively formalised and moreover its physical base changed (e.g. from a driver to a consequence of the eruption, see Section \ref{sect_Introduction}).
Indeed, this conjecture would have to explain how the loop contraction could occur above the stable filament Fil$_2$ located inside an AR, while the eruption of another filament Fil$_1$ occurs at the leading border of the same AR. This requires an analysis, with a numerical simulation, which is outside the scope of present study. 
Still, the observations show that the northern end of filament Fil$_1$ enters slightly within the AR, and reach an end location below L$_3$ loops.  Then, the eruption of Fil$_1$ is clearly involved in this loop contraction while we cannot provide elements in support of the implosion conjecture. {  Even more, the evolution of L$_2$ loops above the stable filament Fil$_2$  and the 
return of the loops to their initial position are not within the framework of the implosion conjecture. }

The loop contraction model of \citet{Zuccarello2017} explained loop contraction by the development of 
magnetohydrodynamic vortexes which develop on the sides of an erupting FR. The loop contraction, then expansion occurs in the field lines overlying the footpoints of the FR.  The numerical simulations 
study the instability of a FR located in a bipolar field simulating a simple and isolated AR. 
This configuration is closer to the magnetic configuration of and above the stable filament Fil$_2$, than of the one of the erupting filament Fil$_1$.  
 Still, following the strongly bended PIL, the contraction/expansion occur on the loops which are next to the northern footpoint of the erupting FR.  Then, the main difference with the simulations of \citet{Zuccarello2017} is that the PIL is so bent that the contracting loops appear on one lateral side, rather than in the continuation, of the erupting filament.  Still, an MHD simulation with the observed multipolar configuration (\fig{mag}) is needed to test the possibility that magnetohydrodynamic vortexes develop and imply the loop contraction/expansion as found before with a simple bipolar magnetic configuration.

%%%%%%%%%%%%%%%%%%%%%%%%%%%%%%%%%%%%%%%%%%%%%%%%%%%%%%%%%%%%%%%%%%%%%%%%%%%%%%%%%%%%%%
\section{Conclusion}
\label{sect_Conclusion}
%%%%%%%%%%%%%%%%%%%%%%%%%%%%%%%%%%%%%%%%%%%%%%%%%%%%%%%%%%%%%%%%%%%%%%%%%%%%%%%%%%%

%   {\S}{\bf --- Summary of observations} \\
 We analyze a filament eruption of 2013 March 16 located close to the west limb. 
The filament, Fil$_1$, was located in between the leading polarity of a decaying active region and a westward remnant dispersed polarity of opposite sign.  In this weak magnetic field environment no GOES flare was reported associated to this eruption. 
Still, we report an arcade of flare loops and two separating J-shaped flare ribbons on the opposite sides of the PIL of the erupting filament.  Later on, the eruption leads to a CME with moderate speed (about 800 \kms ).  All these observational characteristics fit well within the standard 3D model of solar eruptions (\sect{Introduction}).

%   {\S}{\bf --- Dynamics of the erupting filament} \\
The dynamics of the erupting filament is well fitted with a model adding a linear and an exponential 
increases of height with time.   This quantifies two erupting phases: a slow rise and an acceleration 
phases which characterize two different physical mechanisms. A slow rise is indeed expected to occur 
as a consequence of the observed cancelation of the photospheric magnetic field at the PIL.  This process 
is expected to build a FR which is progressively rising in height towards a new equilibrium. Recurrent 
brightenings and restructuration are observed at the northern footpoint of the filament during this phase. 
Next, the exponential growth of height during the acceleration phase characterizes an instability, which 
is likely the torus instability since no significant writhing of the erupting filament is observed 
(as would be present for a kink instability).  The onset of the instability is estimated with the 
time when the exponential growth of the upward velocity becomes significant on top of the previous 
nearly constant velocity.    Another remarkable aspect of the filament eruption was the observations 
of the drift of southern foot-points of the erupting FR, which could be explained by the interchange 
reconnection of the erupting configuration with the  confront magnetic loops of a { neighbouring} active region.
{ Moreover, we also observe the split of the eruptive FR above the southern foot-points with plasma tracing the magnetic connections to two separated magnetic polarities. }
   
%   {\S}{\bf ---  Fil$_2$, loop contraction/expansion } \\
Another filament, Fil$_2$, is present along the internal PIL of the AR.  The northern ends of the filaments are close by, so that both filaments encircle almost the leading polarity of the active region.  
It is then remarkable that Fil$_2$ stays undisturbed while Fil$_1$ is erupting.  
About 21 min after Fil$_1$ eruption onset, the loops L$_3$ suddenly accelerate their contraction 
speed by a factor 2 to 3. At the same time the initially stable southern loops L$_2$,  located above Fil$_2$, contract in a similar way.  Then, the full coronal loop system contract for a period of 22 min. Later on, the loop motion reverses suddenly, then they expand for a period of 20-25 min. The amplitude of the contraction is in between 10 and 20\arcsec . Then, this evolution is well resolved both spatially and temporally by SDO/AIA observations.
Both the contraction and expansion were observed in phase in the full loop arcade on a significant range of projected heights range (about 70 \arcsec ).
Afterwards, these loops set into a stable configuration close to their 
original positions at the start of the fast contraction phase.  
{ The return of the loops to the original position is naturally expected because they pass over the end point of Fil$_1$ and over Fil$_2$, where no or only minimal energy release is expected to occur and the post-eruption equilibrium should be close to the pre-eruption equilibrium.}

%   {\S}{\bf ---  Originality of these observations $\rightarrow$ challenge for simulations } \\
The characteristics of the loop contraction/expansion in this eruption contrast with previous studies.
  First, the loop arcade is located  above a stable filament and extends up to one end of another erupting filament.
  Second, the stable filament is located mostly parallel to the erupting one.
  Third, the stable and erupting filaments are on the same PIL encircling the leading PIL of an active region in decay.
  Finally, the loop evolution starts with a contraction followed by an expansion recovery phase. This half period oscillation with a triangular shape is an original characteristic of present event.
  All these characteristics make this studied event challenging for the models of loop contraction 
 since such magnetic configuration was not  observed or modelled before. In particular, this case is different than the one simulated by \citet{Zuccarello2017} where the eruption occurs in the core of the simulated active region, while in present observations the eruption occur at the leading border of the active region.  
 In summary, numerical simulations are needed with a broader variety of magnetic configuration in order to understand the results of present observations. 
In particular, it is worth studying how strongly the results are depending on the bending of PIL, so on the fully 
3D aspects of the involved magnetic configuration.  Such studies can also allow to separate better the characteristics of the different models proposed so far to interpret the contraction and expansion of coronal loops.

{\bf Acknowledgments}\\
 We recognize the collaborative and open nature of knowledge creation and dissemination, under the control 
of the academic community as expressed by Camille No\^us at \url{http://www.cogitamus.fr/indexen.html}. 
The authors thank the open data policy of the SDO and  STEREO teams.
RC acknowledges the support from Bulgarian Science Fund under Indo-Bulgarian bilateral project, DST/INT/BLR/P-11/2019.
P.D. is supported by CSIR, New Delhi. We thank the reviewers for their constructive comments and suggestions.
%\bibliography{sample631}{}
%\bibliographystyle{apj}

\appendix

\section{Appendix: Earlier Loop Contraction}
\label{sect_Earlier_Contraction}
\setcounter{figure}{0}
%\counterwithin{figure}{section}
%#########################################################
\begin{figure*}[t!]   % Figure 9
\centering
\includegraphics[width=0.9\textwidth]{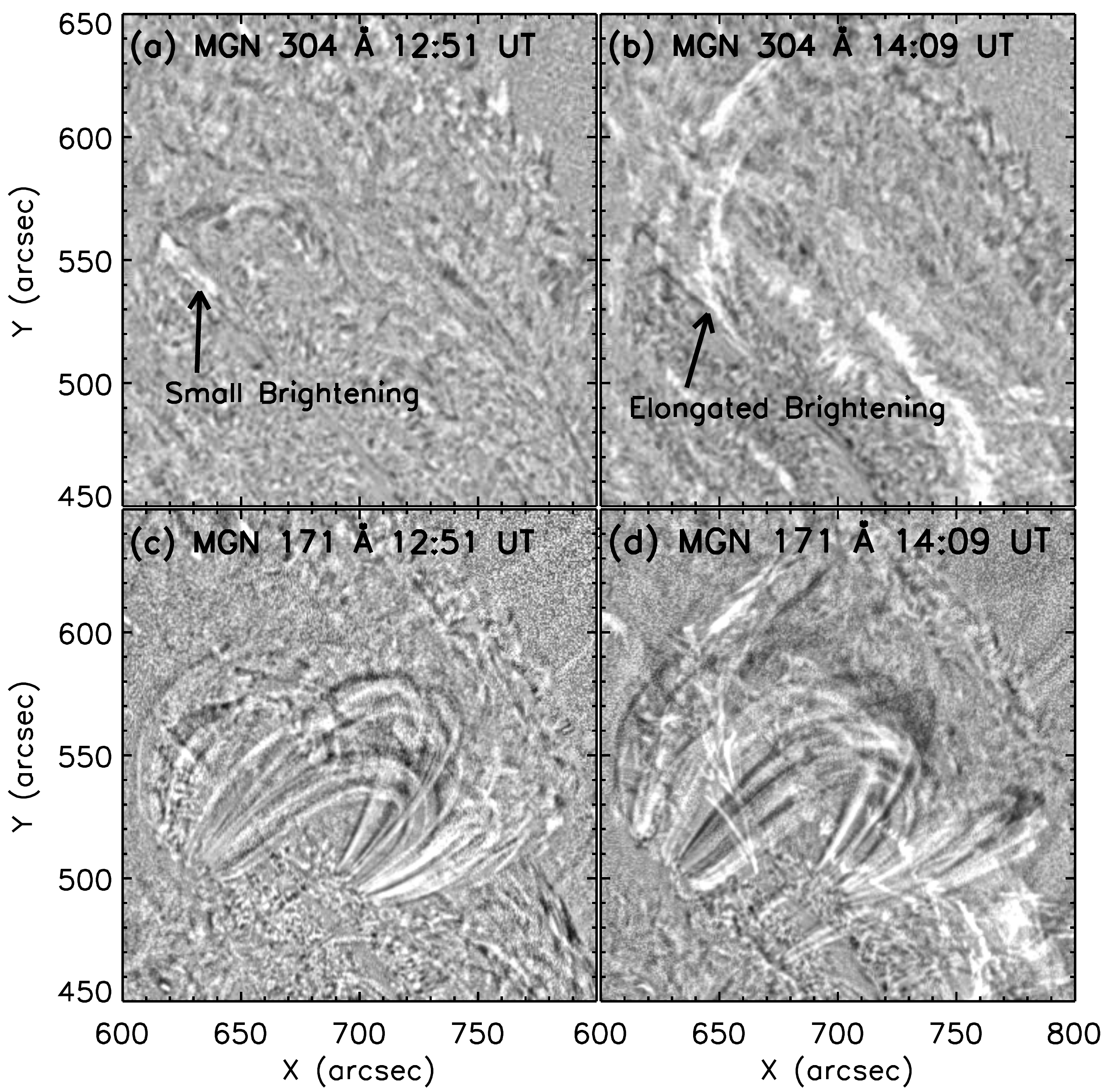}
\caption{ MGN base difference images of AIA 304 and 171 \AA\  at 12:51 and 14:09 UT. The base time for these images 
is 12:30 UT. The small brightening before the eruption and the elongated brightening during the eruption are 
shown in panels (a,c) and (b,d), respectively.  These brightenings are located at the end of filament 
Fil$_1$, close and partly below the L$_3$ loop system (bottom panels). 
The associated movie is available in the Electronic Supplementary Materials. The animation starts at 12:30 UT and end at 15:00 UT. The realtime duration of the animation is 25 seconds.
 } 
\label{fig_bd_171}
\end{figure*}
%#########################################################
   
In order to understand the earlier contraction of loop system L$_3$ (panels (d) of \figs{slice_171}{slice_193}), we studied based difference movies of AIA 304 and 171 \AA\ without and with MGN technique applied before subtracting the image at 12:30 UT.  A main brightening episode is starting at 12:42 UT and is illustrated in the left panels of \fig{bd_171} in the middle of its duration.  This local event is associated with local plasma motions indicating a local restructuration of the magnetic field of the northern end of filament Fil$_1$. 
   A second brighter event started at 13:30 UT in a close by location, while more extended both 
 north, along Fil$_1$, and southward inside the AR (\fig{bd_171}, right panels). It shows 
that, at the beginning of eruption, Fil$_1$ was entering slightly inside the AR (as drawn 
in Figure \ref{fig_mag}d). This event is the activation and reconfiguration 
of Fil$_1$ northern end.  The start of this event is later by at least 11 minutes than 
the eruption onset define in \fig{ht}b (\sect{Kinematics}). 
   These local events are best seen in 304 \AA , while they are also present in 171 \AA\ which 
allows to locate them well with respect to the loops L$_3$ (\fig{bd_171}, bottom panels). 
We conclude that these brightenings are the only indications present in the data of 
magnetic reconfiguration occurring in the vicinity of loops L$_3$ during the linear and 
and early exponential phases of the filament rise.  Then, the data provide no clue about 
the earlier contraction of loop system L$_3$.  This earlier contraction is possibly 
associated to the upward motion of the magnetic configuration of Fil$_1$ in the linear phase. 
 In this case, it has the same physical origin as the contraction present later on in the 
exponential phase (panels (c) and (d) of \figs{slice_171}{slice_193}).

%%%%%%%%%%%%%%%%%%%%%%%%%%%%%%%%%%%%%%%%%%%%%%%%%%%%%%%%
%-------------------------------------------------
\bibliographystyle{aasjournal}
\bibliography{reference}
\end{document}